# Towards Memory Specialization:
# A Case for Long-Term and Short-Term RAM

Peijing Li\*, Muhammad Shahir Abdurraman\*, Rachel Cleaveland\*, Sergey Legtchenko†, Philip Levis\*,
Ioan Stefanovici†, Thierry Tambe\*, David Tennenhouse°, Caroline Trippel\*

\*Stanford University  †Microsoft Research  °Independent Researcher
Stanford, CA, USA  Cambridge, England

**Abstract**
Both SRAM and DRAM have stopped scaling: there is no technical roadmap to reduce their cost (per byte/GB). As a result, memory now dominates system cost. This paper argues for a paradigm shift from today's simple memory hierarchy toward specialized memory architectures that exploit application-specific access patterns. Rather than relying solely on traditional off-chip DRAM and on-chip SRAM, we envisage memory systems equipped with additional types of memory whose performance trade-offs benefit workloads through non-hierarchical optimization.

We propose two new memory classes deserving explicit OS support: long-term RAM (LtRAM) optimized for read-intensive data with long lifetimes, and short-term RAM (StRAM) designed for transient, frequently-accessed data with short lifetimes. We explore underlying device technologies that could implement these classes, including their evolution and their potential integration into current system designs given emerging workload requirements. We identify critical research challenges to realize what we believe is a necessary evolution toward more efficient and scalable computing systems capable of meeting future demands.

## 1 Introduction

For decades, computer system performance improvements have been driven by the continual scaling of memory and compute from Moore's Law. For memory, that scaling has ended – the two dominant memory technologies, SRAM and DRAM, have hit fundamental physical limitations. This has made memory the primary performance, power, and cost bottleneck for current and future computing systems.

These scaling limitations have created a critical divergence between compute and memory capabilities. While processor and network performance continue to improve through architectural advances, SRAM and DRAM densities and costs have stagnated. This divergence has profound economic implications: memory now dominates system cost, with DRAM accounting for over 50% of server hardware costs [34]. As memory to compute capacity ratios fall, applications are becoming increasingly memory-bound.

This challenge has been exacerbated by the rise of memory-intensive workloads, particularly artificial intelligence.
The need for increased bandwidth has driven innovations such as High Bandwidth Memory (HBM), which uses advanced packaging to place many DRAM interfaces close to compute elements [31] for lower latency and higher energy efficiency [33]. However, such packaging improvements alone cannot solve the scaling crisis – they cannot overcome the fundamental scaling limitations of SRAM and DRAM technologies.

Recent research has begun exploring memory technologies that trade off characteristics such as retention[1] and endurance[2] for workload-specific memory access patterns and data lifetimes[3]. For example, Managed Retention Memory (MRM) [24] seeks to repurpose storage-class memories to better serve workloads like large language model (LLM) inference by trading retention time and write performance for improved endurance and lower I/O energy. Similarly, for on-chip memory arrays, gain cell embedded DRAM [28] achieves DRAM-like density with SRAM-like integration properties through similar retention trade-offs.

Building on these insights, we propose a fundamental shift from hierarchical to specialized memory architectures that leverage workload-specific access characteristics. Instead of exposing numerous heterogeneous device technologies directly to system software, we advocate for two new abstractions for classes of memory device technologies: long-term RAM (LtRAM) for persistent, read-heavy data, and short-term RAM (StRAM) for ephemeral, write-intensive workloads. These abstractions provide manageable interfaces for system optimization while enabling diverse underlying implementations tailored to meet the performance requirements of each class.

This paper makes the three following contributions:
1. It proposes LtRAM and StRAM as new memory classes that define specific performance and lifetime characteristics needed by modern workloads.
2. It explores underlying device technologies that could implement LtRAM and StRAM and examines their integration into current system designs.

---
[1]Retention is the time that data is reliably stored in a memory cell without requiring a refresh.
[2]Endurance is the number of write cycles a memory cell can support before it permanently degrades.
[3]Data lifetime is the duration for which the data must remain persistent and accessible in memory hardware.

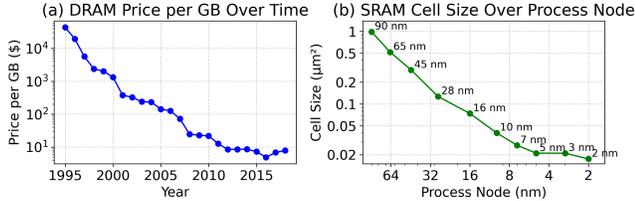

Figure 1: SRAM and DRAM scaling [4]

3. It identifies key research questions that must be resolved to realize the potential of LtRAM and StRAM in both single-node and distributed computing contexts.

## 2 Memory Today: SRAM and DRAM

The typical memory hierarchy in today's computing systems consists of SRAM caches on the compute chip (L1, L2, etc.), off-chip DRAM main memory, and NAND flash or other non-volatile storage for permanent data, as described below in Table 1.

Static RAM (SRAM) consists of six transistors per bit, operates at high speed, statically retains its state while powered, and integrates easily with compute logic. These properties have made SRAM the default choice for on-chip caches and scratchpad memories.

Dynamic RAM (DRAM) consists of a single transistor and capacitor per bit. The capacitor stores the charge representing a 0 or 1, while the transistor gates access to the cell. To maximize density, DRAM capacitors are very narrow and tall, making integration with high-performance logic impractical [18, 36]. DRAM cells have limited retention due to charge leakage and are disrupted by read operations. Therefore, DRAM subsystems dynamically "refresh" by reading and rewriting cell contents (e.g., every 32ms for DDR5). These periodic refreshes consume significant energy even when memory is not actively accessed.

### 2.1 The End of Scaling

Both SRAM and DRAM face the same fundamental challenge: they have stopped scaling. Figure 1(a) shows that DRAM cost trends have stagnated over the past 15 years. While smaller DRAM cells remain technically feasible, manufacturing difficulties from capacitor sizing constraints prevent reductions in per-cell cost [4]. Although 3D stacking can extend density scaling somewhat, the HBM roadmap indicates that this density increase will stop beyond 20 dies due to packaging complexity and cost limitations, placing an upper bound on the capacities of HBM devices [11].

SRAM faces similar constraints. Figure 1(b) shows that SRAM cell sizes have stopped shrinking significantly since the 7nm process node [43]. This stagnation stems from the fundamental physics of SRAM operation, where maintaining the delicate balance between transistor sizing and threshold voltages becomes increasingly difficult at smaller process nodes and lower operating voltages [17].

### 2.2 Packaging Options

One current approach to address the scaling limitations of SRAM and DRAM is to innovate in packaging rather than cell technology.

SRAM is typically integrated inside compute chips but can also be 3D stacked for greater capacity [37]. There are even more packaging options for DRAM. Historically packaged in DIMMs (Dual Inline Memory Modules) for modularity, DRAM packaging increasingly focuses on low power and/or high bandwidth options. LPDDR achieves lower power and higher bandwidth (10Gbps/pin versus 5.6Gbps/pin in conventional DDR) [38] but must be soldered to boards, sacrificing the reconfigurability of DDR DIMMs [26]. HBM achieves much higher bandwidth through 3D stacking of multiple dies connected via silicon interposers [31]. This complex process requires advanced facilities and lowers manufacturing yield, making HBM substantially more expensive than DDR DRAM [34] while constraining capacity and making terabyte-scale on-package configurations impractical [22].

These packaging innovations offer different performance/cost trade-offs but do not solve the fundamental scaling limitations of the underlying SRAM and DRAM cell technologies.

## 3 Escaping the Memory Scaling Trap

To address the scaling challenges of existing memories, a wide range of alternative memory technologies has emerged with fundamentally different scaling roadmaps, including gain cell embedded DRAM [28], ferroelectric RAM (FeRAM) [30], magnetoresistive RAM (MRAM) [9], and resistive RAM (RRAM) [39]. Each technology offers different trade-offs in density, endurance, retention, read/write energies, access times, and on-chip integration capabilities, making them suitable for different applications even though they are not drop-in replacements for SRAM or DRAM. This section quantifies their scaling benefits, examines the constraints that prevent drop-in replacement, and demonstrates how these limitations necessitate a shift toward specialized memory systems.

### 3.1 Density Benefits over SRAM and DRAM

Many of the emerging memory devices are designed to be denser than SRAM or DRAM, while having the potential of maintaining that advantage as process nodes shrink. One such example is gain cell embedded DRAM, which can achieve 2-3x higher density than SRAM at similar process nodes [15] while also allowing on-chip integration due to its all-transistor structure.

More quantitatively, RRAM demonstrates lower read energy and latency than DRAM at higher densities due to its simple cell structure consisting of either one transistor and one resistor or only a single resistor [39]. Scaling trends for RRAM are illustrated in Figure 2, which compares the area densities of DRAM and HBM[4] with various RRAM configurations, from 2D planar RRAM cells [41] to 3D-stacked V-RRAM architectures with 8 and 64 layers [3, 20, 29].

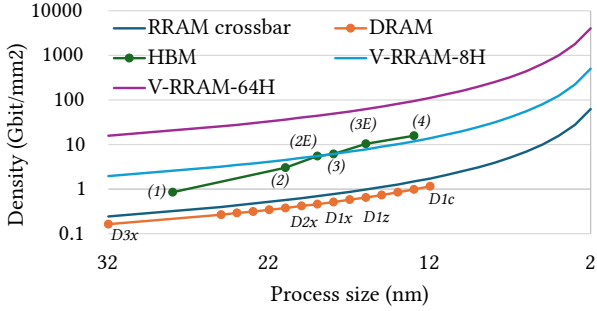

Figure 2: Scaling trends and potentials of DRAM[7] and HBM compared to RRAM.

Figure 2 demonstrates that while planar RRAM cells achieve up to 2x the density of DRAM, the most significant scaling benefits emerge from 3D V-RRAM architectures, which can reach up to 10x the density of state-of-the-art HBM4 at the same process node. These advantages extend beyond current nodes: RRAM can continue scaling to smaller process sizes [42] where DRAM would struggle below 11nm, and V-RRAM architectures can stack up to 64 layers compared to HBM's 16-die limit.

### 3.2 Constraints of Emerging Memory Technologies

While emerging memory technologies offer compelling density advantages, they cannot serve as drop-in replacements for SRAM or DRAM due to fundamental trade-offs in their intrinsic cell properties. Instead, these technologies exhibit asymmetric read/write performance and make explicit trade-offs between retention time and write endurance.

For instance, RRAM's limited endurance and high write energy prevent its use for frequently overwritten transient data [39], despite its superior scaling potential. Similarly, gain cell RAM requires periodic refreshes due to its reliance on intrinsic transistor capacitance rather than SRAM's feedback mechanism for data retention.

These constraints necessitate aligning data access patterns with underlying cell technology characteristics. There are two key properties to consider. First, for memories with limited endurance, the write rate of the data over the expected service life of the device should be consistent with the cell endurance. Second, matching the data retention of the cells to the lifetime of the data helps reduce energy consumption. This motivates memory specialization, where memory systems can be built with a heterogeneous set of memory device arrays, each optimized for specific write patterns and data lifetimes, such that they can cover the full spectrum of memory accesses in a target workload.

---

[4]HBM uses the same process size as DRAM, and stacks 4 dies for HBM1, 8 dies for HBM2, 12 dies for HBM2E and 3, and 16 dies for HBM3E and 4

## 4 A Proposal for LtRAM and StRAM

The broadening landscape of emerging memory technologies poses a significant challenge for system software. Specialized, heterogeneous memory systems that may be unique to individual devices cannot remain entirely transparent to the software stack. Instead, operating systems, middleware, and applications must adapt to different memory devices with varying properties across systems. Moreover, software must remain compatible with both current and future memory devices without requiring constant updates as new specializations emerge.

We propose taming this complexity by organizing emerging devices into just two new memory classes: Short-term RAM (StRAM) and Long-term RAM (LtRAM). As shown in Table 1, these additions create a five-class memory system that preserves existing SRAM, DRAM, and NAND flash (along with other forms of non-volatile storage) while adding targeted capabilities for emerging workloads. Rather than managing numerous individual memory technologies, this approach provides operating systems with manageable abstractions that accommodate diverse underlying device implementations. While one could envision a finer-grained continuum of retention and access characteristics, these two classes capture the majority of emerging device and workload trade-offs without excessive segmentation.

SRAM, DRAM, and NAND flash in Table 1 are already well-established in today's memory hierarchies. We focus the remainder of this section on the two additional memory classes, StRAM and LtRAM, which complement existing technologies to service specific workloads.

### 4.1 Short-term RAM (StRAM)

Short-term RAM (StRAM) is a memory class optimized for transient data that is frequently accessed or read once and discarded, with short sub-second data lifetimes and low energy costs. It is underpinned by memory technologies that have symmetric read/write performance and very high (or unlimited) endurance, but not necessarily high retention. Key use cases of StRAM exploit data transience and include intermediate activation buffers in neural networks, pointer chasing, and temporary data structures such as FIFOs and message queues in server applications.

StRAM can be implemented using various underlying device technologies, with gain-cell embedded DRAM serving as a prime example. Integration approaches range from on-die placement to near-compute configurations, depending on device characteristics and thermal constraints.

### 4.2 Long-term RAM (LtRAM)

Long-term RAM (LtRAM) is a memory class optimized for persistent, read-heavy data with long data lifetimes on the order of minutes and longer, building upon managed-retention concepts [24]. Unlike StRAM, LtRAM prioritizes read performance and energy efficiency over write characteristics, making explicit trade-offs that accept higher write latencies and energy costs, as well as lower endurance, in exchange for superior read properties and retention characteristics. Example technologies that could underpin LtRAM include non-volatile memories such as RRAM, MRAM, FeRAM, and managed-retention DRAM variants. These offer different trade-offs with respect to read/write asymmetry, density, endurance, and integration complexity, and can be deployed on-die, in 3D-stacked packages, or off-chip modules depending on application needs.

| | SRAM | DRAM | NAND | StRAM | LtRAM |
|---|---|---|---|---|---|
| **Strengths** | Low R/W latency, long retention, low static power | Very dense | Extremely dense, low static power | Dense, low write energy, low static power | Very dense, low read energy, low static power |
| **Weaknesses** | Low density | Off-chip only, high R/W energy, high static power, refresh overhead, destructive reads | Off-chip only, low bandwidth, limited endurance, expensive erases | Short retention, refresh overhead | High write energy, high write latency, limited endurance |
| **Uses** | Fast R/W caches | Large, random access, read/write data | Storage, rarely accessed data | Fast R/W caches, Write-and-read scratchpads | Read-mostly data, read-mostly caches |

Table 1: Five Types of Memory. "Power" in this table refers to the continual, leakage draw from the technology, while "energy" refers to the cost of an active read or write operation of a memory cell. Static power only changes with clock frequencies while active energy scales with use.

The key insight motivating LtRAM is that long data lifetimes and read-heavy access patterns allow optimizations that are unsuitable for general-purpose memories. Primary applications include model weights in ML inference, code pages, hot instruction paths, and relatively static data pages —workloads that can tolerate higher write costs in exchange for lower read energy and improved cost per bit. This specialization addresses fundamental mismatches in current systems where read-intensive data competes for the same resources as frequently modified data.

## 5 Specialization Opportunities

Analysis of workload memory access patterns is crucial for identifying specialization opportunities and has precedent in memory systems research. For example, the MRM project [24] examined access patterns in large language model (LLM) inference, finding that the workload is predominantly read-intensive and high bandwidth. This made general-purpose DRAM too slow while rendering HBM over-provisioned in write performance and endurance, leading the authors to propose MRM as a specialized memory class[5] optimized for LLM workloads. Memory access patterns across applications exhibit similar specialization opportunities that extend well beyond LLM inference to encompass the full spectrum of modern computing systems. The following examples demonstrate how application-specific access pattern and data lifetime observations can inform the design of specialized memory systems using the LtRAM and StRAM classifications.

### 5.1 Server Workloads

Server applications exhibit diverse memory access patterns that could benefit from specialized memory technologies.

An interesting example is in-memory data stores such as Redis [6], Memcached [13], and Ray Plasma [1] or in CDNs, DNS and other typical server workloads. They demonstrate read-mostly workloads with variable data lifetimes depending on the exact data caching and access algorithm selected at runtime – we envision that more future applications may adopt this kind of design to suit different compositions of LtRAM and StRAM in particular memory architectures.

Logging, telemetry, and event buffer systems are characterized by write-once workloads with low to medium lifetimes, where data is frequently appended but rarely accessed; StRAM is well-suited for storing the intermediate states of these workloads before the data is eventually archived to off-chip flash or other non-volatile storage. Search engines such as ElasticSearch [23] and Lucene [5] maintain large inverted indices that are read-intensive with long lifetimes, while their query processing generates short-lifetime data structures. Serverless computing platforms create ephemeral execution environments where code and data have extremely short lifetimes, often measured in seconds. Database buffer pools exhibit complex access patterns where recently accessed pages remain hot with shorter data lifetimes while others become cold with long data lifetimes, creating opportunities for heterogeneous memory management between StRAM, LtRAM, SRAM, and DRAM.

Code pages present another compelling case for specialization – they are read-intensive with very long lifetimes, yet current systems store them in the same DRAM as frequently modified application data. Storing them in LtRAM would allow these pages to be stored in a more energy-efficient and cost-effective manner.

---

[5]For our purposes, MRM is a sub-class of LtRAM.

### 5.2 Machine Learning Workloads

AI workloads further motivate the transition towards memory specialization due to their highly predictable and deterministic data flows. These characteristics make the traditional characteristics of random-access memory misaligned with application needs.

During inference, model weights are typically immutable, leading to frequent high-bandwidth large block reads, especially for large language models. As noted in previous work [24], traditional HBM and DRAM are over-provisioned in terms of write performance, hence LtRAM may allow for better efficiency and opportunities for on-chip integration. Activations and intermediate results during inference are immediately discarded after computation, and hence are more suitable for StRAM.

Training workloads combine read- & write-heavy accesses to model weights with write-heavy access to gradients and optimizer states. Activation data during training exhibits particularly distinctive temporal locality – it is intensively accessed during forward and backward passes, then immediately discarded after gradients are computed. The short data lifetimes of activations and gradients make them good candidates for StRAM usage.

### 5.3 Memory Access Patterns within Processor Cores

Within processor cores, memory access patterns similarly mismatch current technology provisioning. Most data persists in SRAM-based L1/L2/L3 caches for only short periods, including function call stacks, local temporaries, intermediate results in math kernels, and tight thread communication. SRAM is over-provisioned for these short-lived data objects, where StRAM would suffice while providing similar performance characteristics at potentially higher density and lower static power consumption.

## 6 Systems Design Challenges

The introduction of LtRAM and StRAM fundamentally disrupts traditional memory system design, creating new research challenges that span the entire system stack. We identify several critical research questions that must be addressed to realize the potential of specialized memory systems.

### 6.1 Abstractions for Post-Hierarchical Memories

Traditional memory systems expose uniform abstractions in byte- or block-addressable, flat address spaces that hide device complexity from software, where the entire memory address space is treated as a homogeneous resource. While this approach provides a simple interface and programming model, it cannot exploit the specialized characteristics of emerging memory technologies or the application-specific access patterns that motivated LtRAM and StRAM.

Traditional hierarchies assume that proximity correlates with performance: SRAM is closest to the processor, fastest, and most expensive; DRAM and NAND flash are progressively further away, slower, and cheaper. Our proposed memory classes underpinned by specialized memory devices break these assumptions from the strict hierarchy, necessitating non-hierarchical optimizations for data placement and access policies.

For example, heterogeneous combinations of SRAM and denser StRAM may both be incorporated on-chip as first-level scratchpad memories, with data placement determined by application requirements [44] rather than hierarchical positioning. Similarly, LtRAM may be placed off-chip yet directly accessed for read-heavy, long-lived data despite the lower bandwidth [45].

Navigating this landscape of "post-hierarchical memories" requires exposing the characteristics of individual memory classes to applications and system software. In literature, NAND flash storage increasingly exposes internal organization to enable more efficient accesses [2], while GPU architectures provide explicit SRAM memory access instructions for performance-critical libraries [14]. These examples demonstrate that providing low-level hardware control can yield considerable performance benefits, albeit with increased code complexity.

The key research challenge becomes: how should operating systems expose retention characteristics, endurance limitations, and read/write performance asymmetries to applications while maintaining programmability? New memory system abstractions must balance performance optimization opportunities with software complexity.

### 6.2 Data Placement Policies

Heterogeneous memory systems require sophisticated data placement policies that determine *when* and *where* placement decisions should be made without relying on hierarchical assumptions.

Current memory management operates at coarse page-level granularities that cannot capture the nuanced patterns that would benefit from memory specialization [27]. Future systems require fine-grained profiling of data lifetimes, read/write rates, and temporal locality patterns [25]. Developing lightweight profiling techniques without significant overhead represents a critical research challenge. Additionally, propagating the insights from profiling remains an open question. We propose that compiler annotations or instruction metadata may provide useful hints about application memory access patterns and requirements.

The timing and authority for placement decisions varies significantly across applications. Specialized software such as databases and ML frameworks possess domain knowledge to make informed placement decisions autonomously given their simplistic dataflows, while general-purpose applications require automatic placement by the OS or hardware. This diversity necessitates flexible placement frameworks that accommodate both explicit application control and transparent system management. However, automated data placement policies in our post-hierarchical memory systems ultimately require prediction of future access pat-terns, where misprediction consequences extend beyond traditional performance penalties, making robust placement algorithms essential for system reliability.

### 6.3 Coherence and Consistency

By disrupting the traditional memory hierarchy, specialized memory classes introduce new challenges for coherence and consistency protocols. Cache coherence ensures consistent data views across processors through invalidation-based protocols (notifying caches when data becomes stale) or update-based protocols (propagating new values directly to all caches). Consistency protocols, on the other hand, ensure that memory operations across different processors appear to execute in a globally agreed-upon order, preserving the illusion of sequential consistency. The previous assumptions of these protocols where all memory types have similar access latencies and read/write costs no longer hold in heterogeneous memory systems.

StRAM's limited retention time introduces novel coherence complexities. Cache controllers must prevent coherence violations by periodically refreshing cache lines before retention limits expire or preemptively evicting them. Alternatively, systems might deliberately allow certain invalidated StRAM cache lines to become stale, trading consistency for reduced refresh overhead.

LtRAM's asymmetric read/write characteristics challenge existing mechanisms. Invalidation-based protocols that rely on frequent metadata updates may perform poorly with LtRAM's high write costs, making update-based protocols more efficient despite higher bandwidth requirements. System designers might also separate cache line data from coherence metadata, storing frequently-updated coherence information in faster memory types while keeping actual data in read-optimized LtRAM.

Cache organizations spanning multiple memory technologies, such as L2 caches implemented across SRAM, StRAM, and LtRAM proposed in [25], must handle variable access latencies of different memory devices while maintaining proper ordering guarantees. This heterogeneity may require entirely new approaches, such as message-passing protocols between memory types, to preserve consistency without sacrificing specialization benefits.

### 6.4 Power-/Thermal-Aware Integration

The sustained growth of cloud computing and now AI have cause significant increases in energy consumption and power densities of data centers. While typical CPU-based compute and storage servers today require 20 kW [21], current-generation AI racks draw as much as 120 kW [16], and next-generation AI racks (slated to arrive late-2027) will draw 600 kW [32]. Today, operators are incapable of running full AI racks in existing installations due to power delivery constraints. Increasingly, operators are turning to on-site power generation [35], even considering nuclear power [40], while evolving power delivery inside the data center [19] for 1+ MW per rack. As such, reducing power

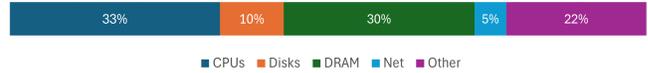

Figure 3: Peak power usage breakdown by components of a warehouse scale computer [10]

consumption is becoming a first-order optimization objective for new systems.

While compute typically receives the most attention in regards to power, memory also makes up a large fraction of overall power draw in modern systems, as shown in Figure 3. Memory contributes significantly to overall system power consumption through static leakage, refresh operations, and data movement costs [8] of both on-chip and off-chip memory arrays.

Memory specialization offers substantial power optimization opportunities through better matching of cell characteristics to workload requirements. At the memory cell level, matching data lifetimes as closely as possible to the cell-level retention characteristics can substantially reduce static power consumption by minimizing unnecessary SRAM leakage and DRAM refresh operations. Data movement energy, which scales super-linearly with interconnect distance, typically dominates total memory power consumption. Thus any attempt at memory specialization must consider the energy consumption of interconnects and packaging in the system. Co-optimization of the memory cells, interconnect, packaging, and data assignment is essential.

Increasing power densities and tight integration of dense memory arrays also means careful attention must be paid to cooling, to ensure effective and reliable operation of the hardware. This might involve potentially inventing and integrating new technologies (e.g., microfluidics [12]) to achieve this.

## 7 Conclusion

As memory scaling stalls, memory has become the primary bottleneck in cost, performance, and power for modern systems. This paper advocates for a shift toward specialized memory architectures, introducing LtRAM and StRAM as abstractions that capture the essential trade-offs of emerging technologies. These classes enable system software to exploit workload-specific access patterns and data lifetimes, improving efficiency and programmability without excessive complexity.

To realize this vision, we call for research in fine-grained workload profiling, lifetime-aware data placement and compiler optimizations, new interfaces that expose heterogeneous memory properties, and efficient hardware implementations of LtRAM and StRAM. Cross-disciplinary collaboration between material scientists, device physicists, circuit designers, computer architects, and systems researchers will be essential to address these multifaceted research problems.